# Van der Waals Magnet based Spin-Valve Devices at Room Temperature

Bing Zhao[1], Roselle Ngaloy[1], Anamul Md. Hoque[1], Bogdan Karpiak[1], Dmitrii Khokhriakov[1], Saroj P. Dash[1*]

[1]Department of Microtechnology and Nanoscience, Chalmers University of Technology,

SE-41296, Göteborg, Sweden

*Corresponding author: Saroj P. Dash, Email: saroj.dash@chalmers.se

**Abstract**

The discovery of van der Waals (vdW) magnets opened up a new paradigm for condensed matter physics and spintronic technologies. However, the operations of active spintronic devices with vdW magnets are so far limited to cryogenic temperatures, inhibiting its broader practical applications. Here, for the first time, we demonstrate room temperature spin-valve devices using vdW itinerant ferromagnet $Fe_5GeTe_2$ in heterostructures with graphene. The tunnel spin polarization of the $Fe_5GeTe_2$/graphene vdW interface is detected to be significantly large ~ 45 % and negative at room temperature. Lateral spin-valve device design enables electrical control of spin signal and realization of basic building blocks for device application such as efficient spin injection, transport, precession, and detection functionalities. Furthermore, measurements with different magnetic orientations provide unique insights into the magnetic anisotropy of $Fe_5GeTe_2$ and its relation with spin polarization and dynamics in the heterostructure. These findings open opportunities for the applications of vdW magnet-based all-2D spintronic devices and integrated spin circuits at ambient temperatures.

**Introduction**

The creation of van der Waals (vdW) heterostructures by combining two-dimensional (2D) quantum materials with complementary properties can allow the discovery of basic new physical phenomena and the development of new device concepts[1]. The discovery of various 2D quantum materials and their heterostructures starting from graphene, insulators, semiconductors, superconductors, and topological materials has revolutionized both fundamental and applied research[2,3]. The most recent addition to this 2D family are magnets, which have offered various advantages over the conventional magnets and open new perspectives in vdW heterostructure designs[4,5]. In addition to the atomically thin and flat 2D nature of magnets, flexibility, gate tunability, strong proximity interactions, and twist angle between the layers can offer a unique degree of freedom and innovative platform for all-2D vdW devices functionalities[4,5].

Recently, several vdW magnets have emerged with the discovery of insulating $Cr_2Ge_2Te_6$[6], semiconducting ($CrI_3$[7], $CrBr_3$[8]), and metallic $Fe_xGeTe_2$[9,10] nature. The insulating 2D magnets are useful for spin-filter tunneling[8,11] and proximity-induced magnetism[12,13], and the metallic magnets can be used as electrodes in magnetic tunnel junctions[14] and magnetic spin-orbit memory devices[15,16] for energy-efficient and ultra-fast spintronic technologies. However, the demonstration of these device operations with vdW magnets is so far limited to cryogenic temperatures. Although room temperature magnetism and proximity effects have been recently reported using 2D magnets[17–19], the lack of active spintronic device operation at room temperature significantly limits its practical application potential[18,20,21]. Furthermore, a lateral spin-valve device with vdW metallic magnets is not realized yet at room temperature, an essential building block for proposed high speed and low power spin-based memory, logic, and neuromorphic computing architectures [22–25].

Here, we demonstrate for the first time robust a room-temperature spin-valve device using metallic vdW itinerant ferromagnet $Fe_5GeTe_2$ in heterostructures with graphene. Significantly large spin polarization of ~45 % could be detected at room temperature due to the successful fabrication of a good vdW interface with graphene. Taking advantage of the lateral spin-valve device design, we probe the unique spin-polarized carrier injection, transport, precession, and detection functionalities with $Fe_5GeTe_2$/graphene heterostructure. The spin-valve signal could be controlled by electric bias, and measurements with different magnetic orientations elucidate the evolution of magnetic anisotropy and spin polarization due to soft magnetic properties in $Fe_5GeTe_2$. Such device operation advances the synergy between spintronics and 2D materials, and is expected to boost practical applications of vdW magnets in all-2D spin-based integrated circuits at room temperature.

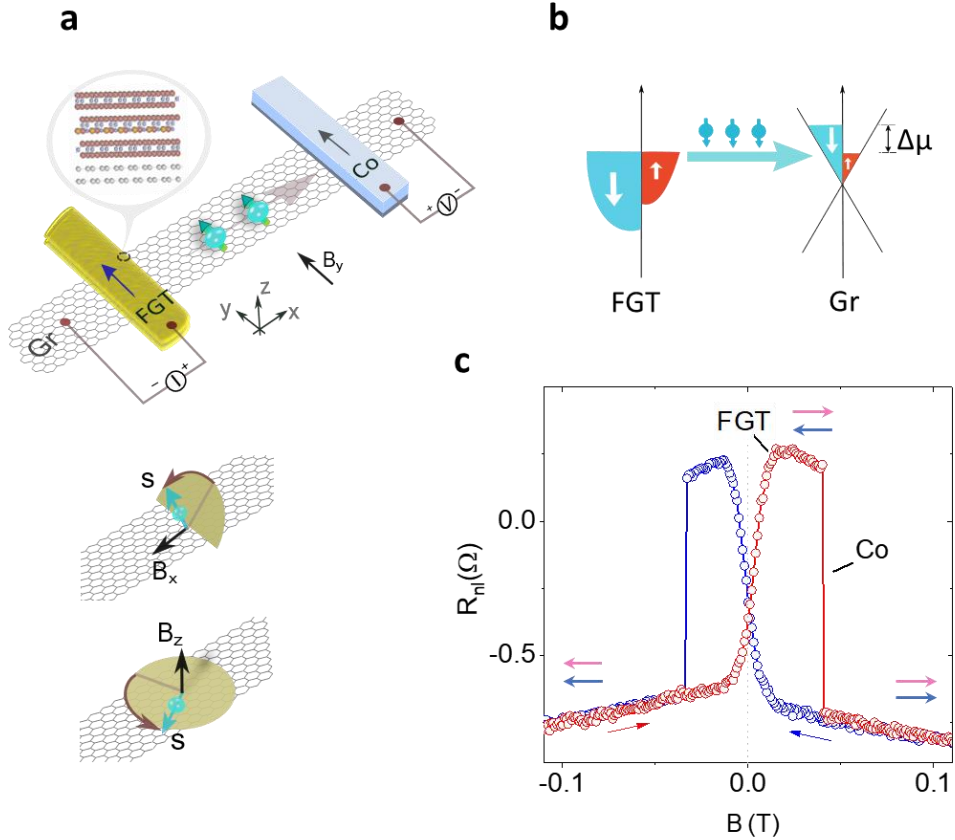

**Figure 1. Room temperature spin-valve device with van der Waals magnet Fe$_5$GeTe$_2$ and graphene heterostructure. a,** Schematic of a spin-valve device with Fe$_5$GeTe$_2$ (FGT) on a graphene (Gr) channel with reference TiO$_2$/Co and Ti/Au (red dots on FGT and graphene) contacts. The top inset shows the schematic of the atomic interface between FGT and graphene. The middle and bottom panels show the schematics for the Hanle spin precession measurements with $B_x$ and $B_z$ fields, respectively. **b,** Schematic illustration for spin injection from FGT into the graphene channel through the vdW gap, inducing a non-equilibrium spin accumulation Δμ in graphene. **c,** The measured nonlocal (NL) spin-valve signal $R_{nl}=V_{nl}/I_{dc}$ for parallel and antiparallel alignment of FGT and Co electrodes (light pink and blue arrows) with an applied bias current $I_{dc}=+15$ μA across the FGT/Gr junction and NL voltage $V_{nl}$ measured by the TiO$_2$/Co detector, as shown in the measurement geometry in a**.** A representative spin-valve data is shown for an in-plane magnetic field sweep at an angle of -60° in the x-y plane relative to the +y axis**.** The arrows show the sweeping direction of the magnetic field. The switching fields for FGT and Co electrodes in the spin-valve measurement are indicated. The measurements were performed in Dev 1 at room temperature.

## Results and Discussion

The schematic of the spin-valve device with measurement geometry is shown in Fig. 1a, consisting of multilayer Fe$_5$GeTe$_2$ as a spin injector (source) or detector (drain) on the few-layer graphene spin transport channel. The measurements with magnetic field (*B*) sweep along the y-axis provide information about in-plane spin polarization of Fe$_5$GeTe$_2$ along the *y*-axis, whereas the Hanle spin precession measurements with B along x(z)-axis is an unambiguous and reliable approach to extract the (initial) injected spin states and its dynamic properties. The vdW gap between $Fe_5GeTe_2$ and graphene should allow an efficient spin injection and detection process in the heterostructure (Fig. 1b). The motivation behind the use of $Fe_5GeTe_2$ is its ferromagnetic order with Curie temperature ($T_c$) close to room temperature[18,21], compared to $Fe_3GeTe_2$ with $T_c$ ~ 220 K[9,10]. An increased saturation magnetization in $Fe_5GeTe_2$ can also provide larger spin polarization due to the Fe-rich vdW heterostructure forming a three-dimension (3D)-like spin-pair interaction[20,21]. The motivation behind the use of graphene as the channel material is its excellent spin transport properties[26] and ideal combination with $Fe_5GeTe_2$ for vdW heterostructure-based spin-valve devices operating at room temperature. The $Fe_5GeTe_2$/graphene heterostructures are prepared on $Si/SiO_2$ substrate employing exfoliation and dry transfer techniques. The nonmagnetic Ti/Au and ferromagnetic ($TiO_2$/Co) contacts are nanofabricated on $Fe_5GeTe_2$ flakes and graphene channels for reference electrodes. The contact resistance of $Fe_5GeTe_2$/graphene is ~ 6 kΩ and $TiO_2$/Co is ~ 1-5 kΩ. For details of the device fabrication, see the Methods section and Supplementary Fig. S1.

**Room temperature spin valve with $Fe_5GeTe_2$/graphene heterostructure**

We first measured the spin-valve operation with spin injection from $Fe_5GeTe_2$ into the graphene channel at room temperature. A source current was applied between $Fe_5GeTe_2$ and the reference Au electrode on graphene, and the nonlocal (NL) voltage was measured between detector Co contact and another reference Au contact (Fig. 1a). Figure 1c shows the measured spin-valve signal with switchings originating from $Fe_5GeTe_2$ injector and Co detector contacts with different coercive fields ($H_c$) for an in-plane magnetic (B) field sweep. By comparing the reference spin valve signal with all-Co injector and detector contacts (see Supplementary Fig. S2a), we can confirm that the sharp switching at higher $H_c$ in the signal is the contribution from the detector Co contact, and the slow and lower $H_c$ switching is originating from the $Fe_5GeTe_2$ injector electrode. The observation of the robust spin-valve signal shows the presence of an in-plane spin component in $Fe_5GeTe_2$ at room temperature.

To investigate the anisotropy in spin injection from $Fe_5GeTe_2$ and to probe the out-of-plane spin component (along the z-axis), an angle dependence (Fig. 2a) of the spin-valve measurement was performed from +90º to -90º in the x-y plane (Fig. 2b and Fig. 2c). The measurements show that the $Fe_5GeTe_2$ has a very soft magnetization and can be easily rotated in the sample plane by a small applied field; however, it has a robust out-of-plane remnant spin polarization. Specifically, several features can be observed from in-plane angle-dependent measurements. First, the switching field of detector Co increases with the magnetic field rotation angle Φ, consistent with the strong uniaxial magnetic anisotropy of Co in a narrow stripe geometry (see Supplementary Fig. S2d and Fig. S2e). Second, the spin-valve signal can be seen as the result of the non-zero in-plane projection of the injected spin polarization from Fe$_5$GeTe$_2$ on the magnetic moment direction of the detector Co electrode. Here, if we suppose the magnetic moment of Co remains along the y-axis in the small field range (-0.1 to 0.1 Tesla, T), we observe that the saturation field of Fe$_5$GeTe$_2$ is smallest at around Φ≈-50°. These measurements show the presence of an in-plane magnetic anisotropy in the Fe$_5$GeTe$_2$ flake.

Third, the ±90° data represents the x-axis Hanle (xHanle) curves for spin precession measurement. Contrary to conventional symmetric xHanle signal with both Co-Co injector-detector devices (see Supplementary Fig. S2d), a sine-shaped signal is observed in the FGT-Co injector-detector geometry, which shows an out-of-plane spin injection $S_z$ from Fe$_5$GeTe$_2$ into the graphene channel (Fig. 2d and Fig. 2e)[27,28]. Notably, there is almost no symmetric Hanle component at Φ=±90°, suggesting minimum remnant spins $S_y$, i.e., no remnant magnetic moment $M_y$ along y-axis when $B_x$=0T. Such observation indicates the soft magnetic property of the Fe$_5$GeTe$_2$ flake with the in-plane component along the x-axis at remanence, however, with the presence of a robust out-of-plane spin component $S_z$. Forth, both the spin valve and the xHanle signals coexist, for example, at Φ=±60° and ±80°. Here, the contribution of $S_y$ projection is identical; however, the $S_z$ spin precession direction is opposite for positive and negative angles Φ, which induces a different sign of the sine-shaped Hanle signal. To extract the sine-shaped component from the nonlocal signals, we analyzed the measured data by removing the Co switching and symmetrizing the switched curves (see Supplementary Note 1 and Fig. S3). The magnitude of the normalized signal to the one at Φ=0° as a function of Φ is presented in Fig. 2f, agreeing well with the cos(Φ) ~ Φ curve. It confirms the projection of in-plane magnetic moment $M_{eff}$ in Fe$_5$GeTe$_2$ on the y-axis versus in-plane magnetic field B rotation follows the cos(Φ) relation. From these observations of spin-valve and xHanle measurements, we found the Fe$_5$GeTe$_2$ flake to be a very soft ferromagnet with both in-plane and out-of-plane magnetic moments.

**Large spin polarization in Fe$_5$GeTe$_2$ /graphene interface**

To calculate the spin polarization of Fe$_5$GeTe$_2$, we consider the magnitude of the spin-valve signal (for Φ=0°), ΔR$_{nl}$=P$_{FGT}$.P$_{Co}$.λ$_{gr}$.R$_{sq}$.exp(-L$_{ch}$/λ$_{gr}$)/w$_{gr}$, where only the spin polarization of Fe$_5$GeTe$_2$, P$_{FGT}$, is unknown; and the other parameters (P$_{Co}$ the spin polarization of Co, λ$_{gr}$ spin diffusion length, R$_{sq}$ square resistance, L$_{ch}$ channel length and w$_{gr}$ width of the graphene channel) are extracted from the reference Co$_1$-Co$_2$ spin valve with the standard Hanle formula. From the standard Hanle data fitting, we extract the spin lifetime in the graphene channel $\tau_s = 198 \pm 13$ ps, spin diffsusion constant D$_s$=0.008±0.001m$^2$/s and spin diffusion legnth $\lambda_{gr} = 1.26\ \mu m$. The spin polarization of Co is P$_{co}$=10.1%±4.1% (see more details in Supplementary Note 4). From the FGT-Co spin valve signal, we calculate the lower limit of effective in-plane spin polarization of Fe$_5$GeTe$_2$, |P$_{FGT}$|=35.8%±1.3%. To be noted, in the spin valve signal, the switching of Co occurs before full saturation of Fe$_5$GeTe$_2$ (for Φ=0°), which is much larger than the Co coercive field. Therefore, a much larger spin valve signal can be achieved if the detector Co contact is designed to have a higher coercive field, and effective |$P_{FGT}$| can be as large as ~44.9% if the fully saturated $Fe_5GeTe_2$ is considered. The out-of-plane spin polarization along the z-axis is calculated to be $P_z$=9.5%±0.3% from x-Hanle data in Fig. 2e. Moreover, the out-of-plane magnetic moment $M_z$ seems to be robust enough in the in-plane x-Hanle measurements and total spin polarization can be calculated by $P_{tot}=\sqrt{(P_{in}^2+P_z^2)}=\sqrt{0.449^2+0.095^2}=45.9\%$ (see Supplementary Note 2 for details). Such a large spin polarization at the $Fe_5GeTe_2$/Gr interface suggests an efficient spin injection efficiency due to the vdW gap, the sharp interface between $Fe_5GeTe_2$/Gr, and the large saturation magnetization of $Fe_5GeTe_2$ at room temperature.

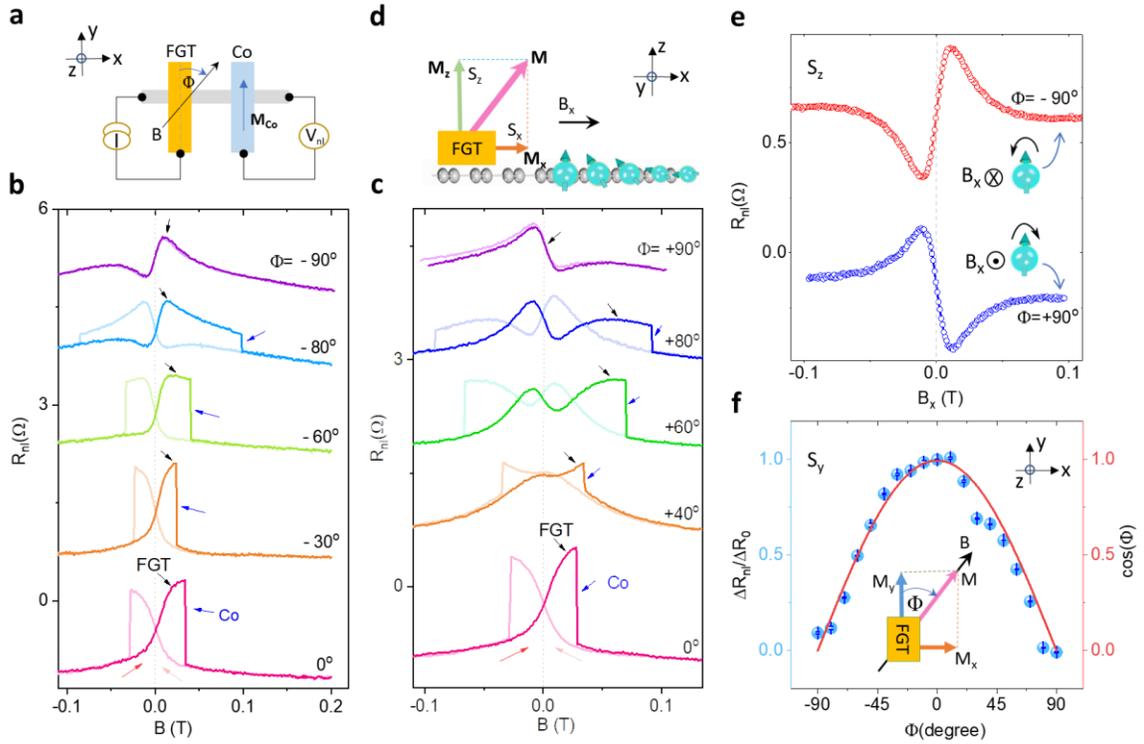

**Figure 2. Anisotropy in spin-valve signal and spin precession. a.** The top view schematic of the spin valve device with in-plane B field sweep in the x-y plane with an angle Φ with respect to y-axis. **b, c.** Angle dependence of the spin valve signal with angle Φ =0 to -90º and 0 to +90º, respectively. The switchings of injector FGT and detector Co electrodes in the spin valve signal are shown by black and blue arrows, respectively. A y-axis shift is added in the signals for clarity, and a small hysteresis of ~3.5mT from the magnet is corrected using control experiments of reference graphene Hanle signals. **d.** The side view schematic of the device for magnetic moment configuration in FGT with an external $B_x$ field. M is the magnetic moment with components $M_x$ ($M_z$) along the x (z)- axis. $S_z$ and $S_x$ are the corresponding injected spin orientations. **e.** xHanle signals at Φ =±90º with parabolic background subtraction. The inset shows the cartoon for spin precession direction with opposite B fields. **f.** Normalized spin valve signal components and *comparison with the cosine function with the in-plane B rotation angle Φ. The error bars are within the data points, calculated based on the noise level of the signal. The inset is the schematic for the magnetic moment projection on the axes, where $S_y$ is injected spins from $M_y$ magnetization of FGT. The measurements were performed in Dev 1 at room temperature.*

**Negative spin polarization of Fe$_5$GeTe$_2$/graphene interface**

To examine the sign of spin polarization of Fe$_5$GeTe$_2$ contacts on graphene, we performed detailed bias dependence measurements and compared them with the standard Co contacts. We investigated the bias current dependence of the spin injection and extraction signal from Fe$_5$GeTe$_2$ on the graphene channel (Fig. 3a), where the direction of the generated spins *s* are found to be dependent on the polarity of the applied current bias. Reversing the bias current direction ($+/-I_{dc}$) in Fe$_5$GeTe$_2$ results in an accumulation of opposite spin polarization in graphene and hence an inverted behavior of the measured spin-valve signal (Fig. 3b). These measurements demonstrate that spin polarization direction can be controlled by electrical spin injection and extraction in Fe$_5$GeTe$_2$. As a reference, the control experiment with the Co$_1$-Co$_2$ spin-valve was also performed with $+/-I_{dc}$ bias currents (Fig. 3c), showing the expected sign change of the signal. The magnitude of the spin-valve signals with bias current for FGT-Co$_2$ and Co$_1$-Co$_2$ spin valves are shown in Fig. 3d for the spin injection and extraction regimes. The accumulated spin density is observed to scale linearly with the applied bias current for both Fe$_5$GeTe$_2$ and Co contacts on the graphene channel.

Surprisingly, we notice that the sign of spin signal with one Fe$_5$GeTe$_2$ electrode is the opposite compared to that of the standard all-Co spin valves, which implies the negative or opposite spin polarization of Fe$_5$GeTe$_2$ compared to Co contacts (see Supplementary Note 3). For example, in Fig. 3e, we plotted the comparison of spin valve signals with FGT-Co$_2$ and Co$_1$-Co$_2$ injector-detector contact configurations for positive bias currents, showing opposite spin polarization between Co and Fe$_5$GeTe$_2$ contacts. Such a negative spin polarization is found to be robust and could be reproducibly observed in multiple devices (also see data from Dev 2 and Dev 3 in Supplementary Fig. S5a and Fig. S5b). A possible origin can be due to the unique band structure with the 'down' spin state as the majority in Fe$_5$GeTe$_2$ at the interface with graphene. These observations agree with a recent theoretical prediction[29], where the Fe atoms, with d-orbital electrons, dominate the contribution of the unique asymmetry in the electron population between spin-up and spin-down states.

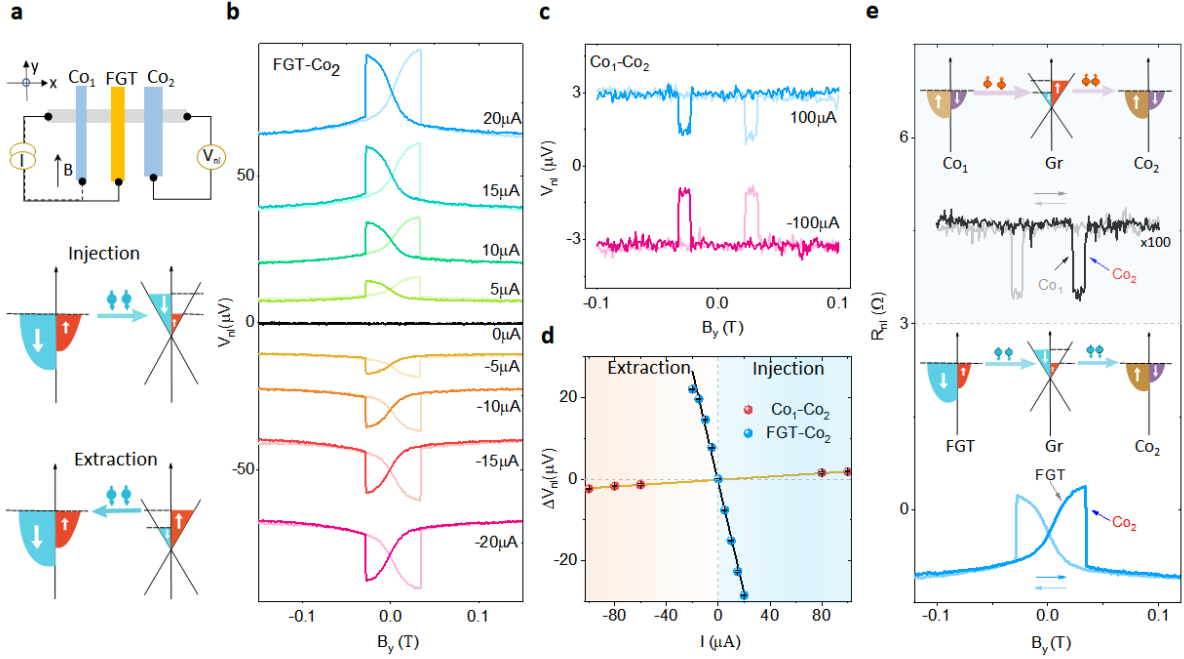

**Figure 3. Negative spin polarization of Fe$_5$GeTe$_2$/graphene interface. a.** Top panel: Schematic of the nonlocal measurement geometries for FGT-Co$_2$ and Co$_1$-Co$_2$ spin valves. Bottom panels: Schematics for the spin injection and extraction process with FGT, showing the accumulation of down and up spins in the graphene channel, respectively. **b.** Spin valve signals with FGT-Co$_2$ for different bias currents of +/-$I_{dc}$. **c.** Co$_1$-Co$_2$ Spin valve signals with $I_{dc}$=±100 µA. **d.** Bias dependence of the spin valve signal magnitude and the linear fittings. The error bars (within the data points) are calculated from the standard deviation of the measured background noise in the data. **e.** Comparison of spin valve signals Δ$R_{nl}$ (=Δ$V_{nl}$/I) with Co$_1$-Co$_2$ and FGT-Co$_2$ injector-detector contact configurations, showing opposite spin polarization between Co and FGT contacts. The Co$_1$-Co$_2$ spin valve signal is multiplied by 100 along the y-axis for clarity. The inset in the top and bottom panels are the schematics of the spin injection and detection process by Co$_1$ and FGT, respectively. Both nonlocal spin valves share the same detector Co$_2$ with the same switching field in the signals. All the measurements were performed on Dev 1 at room temperature.

## Out of plane spin precession and dynamics in Fe$_5$GeTe$_2$/graphene spin valve

Next, to investigate Hanle spin precession of in-plane spin polarization, we carried out the measurements with an out-of-plane magnetic field ($B_z$) sweep[30]. However, with a soft magnet Fe$_5$GeTe$_2$ as an injector, the magnetic moment $M$ and the injected spin $s$ rotate to the out-of-plane direction with a small external $B_z$ field (Fig. 4a). The magnetic moment $M$ also has an in-plane component $M_{eff}$, which induces the spin polarization $S_{xy}$, corresponding to the components along the $y$- and $x$-axis ($S_y$ and $S_x$). As shown in Fig. 4b, we observed a modified Hanle signal due to the combined effect of magnetic moment rotation of Fe$_5$GeTe$_2$ and precession of the injected spins in the graphene channel. Furthermore, measurements with the different magnetization direction of $Fe_5GeTe_2$ and Co electrodes result in the sign reversal of Hanle curves (+/-$M_{eff}$), as shown in Fig. 4c (bottom panel). To eliminate the non-spin-related background from the measured data, we take an average of the signals $R_{avg}$ and decompose them into symmetric (Sys) and antisymmetric (Ays) components (Fig. 4c, top and middle panels), corresponding to the spin components $S_y$ and $S_x$, respectively.

First, considering the symmetric signal for the entirely out-of-plane moment of $Fe_5GeTe_2$ for $|B_z|>0.15$ T, there is no contribution to the spin precession signal and the magnitude of $R_{nl}$ should be null. Notably, we observe that the symmetric signal shows a comparable magnitude of signals at $B_z$=0T and $|B_z|>0.15$ T. Therefore, the in-plane spin $S_y$ at $B_z$~0 T is also negligible and so is the remanent magnetization $M_y$. At the intermediate stages, where $B_z$ is in the range of 0 to ±0.15 T, the $M_y$ increases to a peak value and decreases again to zero for the magnetic moment $M$ saturation along the z-axis (Fig. 4a). This is supported by our simple simulation, which shows that $M_y$ should be minimum at B~0 T (see Supplementary Note 4 and Note 5). Secondly, the antisymmetric component of the signal suggests the existence of the spins $S_x$ along the x-axis. We can fit the $S_x$ component and obtain the effective spin polarization $P_x$=15.3%±4.8% (see Supplementary Fig. S4c.). These observations suggest that the in-plane magnetic moment $M_{eff}$ does not align with the long axis of the flake, which can be due to the in-plane magnetic anisotropy and crystal symmetry of $Fe_5GeTe_2$. Therefore, the observed z-axis Hanle signals are combined effects of magnetic moment rotation of $Fe_5GeTe_2$, and precession and dephasing of the injected spins $S_x$ and $S_y$ in the graphene channel (see more discussions in Supplementary Note 1).

Furthermore, we extracted the magnetic moment rotation angle $\alpha_M$ with $B_z$ from the z-axis Hanle curves using opposite magnetic configurations +/-$M_{eff}$ in Fig. 4c (see details in Supplementary Note 4). It shows the evolution of local magnetic moment near the $Fe_5GeTe_2$/graphene interface with $B_z$ and its saturation at ~ 0.15 T (Fig. 4d). To correlate the orientation of spin injection signal to the magnetization behavior of the $Fe_5GeTe_2$ with the external $B_z$ field, the anomalous Hall effect (AHE) measurement was also performed in the same $Fe_5GeTe_2$ flake (Fig. 4e). The AHE curve suggests a rotation of the magnetization in $Fe_5GeTe_2$ along the z-axis, and the saturation point of the AHE curve is also consistent with that of the $\alpha_M$ ~ $B_z$ relation. Detailed angle dependence of the AHE measurement performed in a similar $Fe_5GeTe_2$ flake (see Supplementary Note 6) suggests a presence of the magnetic anisotropy in $Fe_5GeTe_2$ (see Supplementary Note 2). This observation agrees well with the spin-valve Hanle spin precession results.

In addition to spin injection from $Fe_5GeTe_2$ into the graphene channel, we also demonstrated spin detection by $Fe_5GeTe_2$ using Dev 2 as shown in Supplementary Note 7 and Supplementary Fig. S8. The bias dependence of the spin signals with the $Fe_5GeTe_2$ detector also showed good linearity in the measured bias range. Furthermore, the gate voltage dependence of the spin valve signal for the spin injection and detection by $Fe_5GeTe_2$ shows a tunability (Supplementary Fig. S8c and S9c), mainly due to the modulation of the graphene channel resistance with gate voltage and conductivity mismatch between the $Fe_5GeTe_2$ and graphene channel[31].

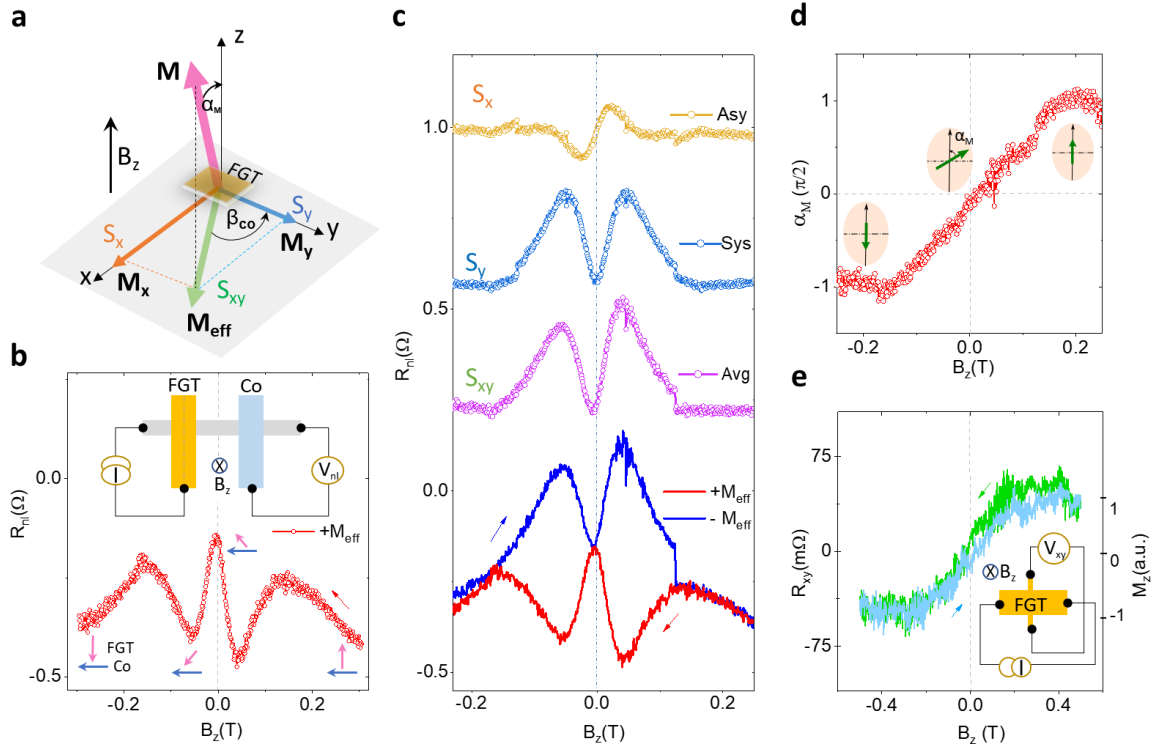

*Figure 4. Out-of-plane Hanle signal in $Fe_5GeTe_2$/graphene spin valve. a.* *Schematic of the magnetic moment configuration in FGT, where M represents the total magnetic moment of FGT, rotating with external out-of-plane magnetic field $B_z$. $M_{eff}$ is the effective in-plane projection of M with components ($M_x$, $M_y$) along the x-and y-axis. $S_{xy}$ is the injected in-plane spins from $M_{eff}$ with components $S_x$ and $S_y$, along the x- and y-axis, respectively. $β_{co}$ is the relative angle between $M_{eff}$ and the y-axis. $α_M$ is the angle between M and the z-axis.* **b.** *Measured Hanle spin precession signal with $B_z$ sweep for $I_{dc}$ = +15 μA with in-plane magnetization +$M_{eff}$ in FGT. The blue and green arrows show the FGT and Co magnetic moment evolution with $B_z$. The red arrow shows the B sweeping direction. The top inset shows the schematic of measurement geometry.* **c.** *Hanle curves for the parallel and antiparallel magnetization direction of injector FGT and detector Co electrodes. Symmetric (Sys) and anti-symmetric (Ays) components of the averaged Hanle signal $R_{avg}$, where $R_{avg}=[R_{nl}(+M_{eff})-R_{nl}(-M_{eff})]/2$, and $R_{ays}=[R_{avg}(B)+R_{avg}(-B)]/2$; $R_{sys}=[R_{avg}(B)-R_{avg}(-B)]/2$.* **d.** *Relation of the extracted $α_M$ as a function of $B_z$. The insets are the corresponding magnetic moment evolution in FGT with the external field $B_z$.* **e.** *Anomalous Hall effect (AHE) signal of FGT ($R_{xy}$) as a function of magnetic field $B_z$ sweeps along the z-axis. The $M_z$ on the right axis is the out-of-plane magnetic moment in an arbitrary unit, $M_z=±1$ suggesting a total out-of-plane magnetic moment saturation. The inset shows the AHE measurement geometry. All the measurements were performed on Dev 1 at room temperature.*

**Summary and Outlook**

In conclusion, we demonstrated a robust room-temperature operation of van der Waals magnet-based spin-valve devices. Highly efficient spin injection, transport, dynamics, and detection could be observed with a large negative spin polarization of $Fe_5GeTe_2$ ~ 45% in the heterostructure with graphene channel. Furthermore, soft ferromagnetic properties and spin polarization anisotropy in $Fe_5GeTe_2$ could be probed via angle dependence measurements in the spin-valve devices. These studies provide unique insights into the room-temperature magnetism of $Fe_5GeTe_2$ and its relation with spin transport and precession in graphene heterostructure devices. These results establish the integration of vdW magnets with graphene spin valve devices and have a vast potential in developing all-2D spintronics devices and integrated circuits at room temperature. Furthermore, this will also bring a strong synergy between 2D quantum materials and spintronics with the possibility of further control the figure of merits by twist angle between the layers, magnetic proximity effects, and gate tunability for energy-efficient and ultra-fast spintronic devices. These room-temperature developments in van der Waals heterostructure will open opportunities for the use of van der Waal magnets for fundamental studies in condensed matter physics and device applications in spintronic sensors, memory, logic, and neuromorphic computing architectures [32,33].

**Methods**

**Fabrication of devices and electrical measurements**

The $Fe_5GeTe_2$ (FGT) crystal (from Hq Graphene) flakes (20~50 nm) were exfoliated and dry transferred onto the exfoliated few-layer graphene (from HOPG) on an $n^{++}$doped Si substrate with 285 nm $SiO_2$. For the device fabrication, the nonmagnetic and magnetic contacts were prepared by multiple electron beam lithography (EBL) processes and electron beam evaporation of metals. The nonmagnetic Ti/Au contacts were first prepared on FGT flakes and graphene for reference electrodes. The ferromagnetic contacts ($TiO_2$/Co) on graphene were prepared in a two-step deposition and an oxidation process was adopted, 0.4 nm Ti was deposited, followed by a 10 Torr $O_2$ oxidation for 10 min each, followed by 60 nm of Co deposition. Measurements were performed at room temperature with a magnetic field up to 0.8 Tesla and a sample rotation stage in vacuum conditions. The electronic measurements were carried out using current source Keithley 6221, nanometer 2182A, and dual-channel source meter Keithley 2612B.

**Acknowledgments**

The authors acknowledge financial supports from EU Graphene Flagship (Core 3, No. 881603), Swedish Research Council VR project grants (No. 2016–03658), 2D TECH VINNOVA competence center (No. 2019-00068), Graphene center, EI Nano, and AoA Materials program at Chalmers University of Technology. We acknowledge the help of staff at Quantum Device Physics laboratory and Nanofabrication laboratory in our Department of Microtechnology and Nanoscience at Chalmers University of Technology.

## Data availability

The data that support the findings of this study are available from the corresponding authors on a reasonable request.

## Author information

### Affiliations

Department of Microtechnology and Nanoscience, Chalmers University of Technology, SE-41296, Göteborg, Sweden

Bing Zhao, Roselle Ngaloy, Anamul Md. Hoque , Bogdan Karpiak, Dmitrii Khokhriakov, & Saroj P. Dash

### Contributions

B.Z and R.N fabricated and characterized the devices. A.M.H, B.K, D.K, S.P.D. participated in device preparation and measurements. B.Z. and S.P.D conceived the idea and designed the experiments. B.Z. and S.P.D. analyzed and interpreted the experimental data, compiled the figures, and wrote the manuscript with inputs from all co-authors. S.P.D. supervised the research project.

### Corresponding author

Correspondence to Saroj P. Dash, saroj.dash@chalmers.se

### Competing interests

The authors declare no competing interests.